\documentclass[showpacs,preprintnumbers,english,amsmath,amssymb]{revtex4}
\usepackage[T1]{fontenc}
\usepackage[latin1]{inputenc}
\usepackage{graphicx}
\usepackage{amssymb}
\usepackage{verbatim}
\usepackage{epic,eepic}
\usepackage{babel}
\usepackage{color}

\begin{document}

\title{Coulomb corrections to density and temperature in heavy ion collisions}

\author{Hua Zheng$^{a,b)}$\footnote{Electronic address: zhengh@tamu.edu}, Gianluca Giuliani$^{a)}$ and Aldo Bonasera$^{a,c)}$}
\affiliation{
a)Cyclotron Institute, Texas A\&M University, College Station, TX 77843, USA;\\
b)Physics Department, Texas A\&M University, College Station, TX 77843, USA;\\
c)Laboratori Nazionali del Sud, INFN, via Santa Sofia, 62, 95123 Catania, Italy.}




\begin{abstract}
A recently proposed method, based on quadrupole and multiplicity fluctuations in heavy ion collisions, is modified in order to take into account distortions due to the Coulomb field. The classical and quantum limits for fermions are discussed. In the classical case we find that the temperature determined from ${}^3He$ and ${}^3H$, after the Coulomb correction, are very similar to those obtained from neutrons within the Constrained Molecular Dynamics (CoMD) approach. In the quantum case, the proton temperature becomes very similar to neutron's, while densities are not sensitive to the Coulomb corrections.

\end{abstract}


\maketitle

\section{Introduction}
Important informations about the Nuclear Equation of State (NEOS) can be obtained by colliding heavy ions \cite{aldo1}. The task is not easy since we have to deal with a microscopic dynamical system. Non-equilibrium effects might be dominant and we have to derive quantities, such as density, temperature and pressure to constrain the NEOS. Recently we have proposed a method to determine density and temperature from fluctuations \cite{sara, hua1, hua2, hua3}. The reason for looking at fluctuations, especially in the perpendicular directions to the beam axis, is because they are directly connected to temperature for instance through the fluctuation-dissipation theorem \cite{landau}.  Of course, the system might be chaotic but not ergodic, but fluctuations should give the closest possible determination of the 'temperature' reached during the collisions. Quadrupole fluctuations (QF) \cite{sara} can be easily linked to the temperature in the classical limit. Of course, if the system is classical and ergodic, the temperature determined from QF and, say from the slope of the kinetic distribution of the particles should be the same. In the ergodic case, the temperature determined from isotopic double ratios \cite{albergo} should also give the same result.  This is, however, not always observed, which implies that the system is not ergodic, nor classical. We can go beyond the classical approximation \cite{hua1, hua2} since we are dealing with fermions. In such a case it is not possible to disentangle the 'temperature' from the Fermi energy, thus the density \cite{hua1}. Because we have two unknowns, we need another observable, which depends on the same physical quantities. In \cite{hua1, hua2, hua3} we have proposed to look at multiplicity fluctuations (MF) which, similarly to QF, depends on $T$ and $\rho$ of the system in a way typical of fermions \cite{hua1, hua2} or bosons,  such as alpha-particles \cite{hua3}. The application of these ideas in experiments has produced interesting results such as the sensitivity of the temperature from the symmetry energy \cite{alan1}, fermion quenching \cite{brian} and the critical $T$ and $\rho$ in asymmetric matter \cite{justin}. Very surprisingly, the method based on quantum fluctuations \cite{justin} gives values of $T$ and $\rho$ very similar to those obtained using the double ratio method and coalescence \cite{qin} and gives a good determination of the critical exponent $\beta$. This stresses the question on why sometimes different methods give different values, including different particles ratios \cite{tsang1, tsang2}, while in other cases the same values are obtained. In \cite{sara} the classical temperature derived from QF gave different values for different isotopes. Clearly the Coulomb repulsion of different charged particles can distort the value of the temperature obtained from QF, which depends on kinetic values.  On the other hand, MF for different particles seem to be independent on Coulomb effects as we will discuss below \cite{hua3}. Also the obtained values, say of the critical temperature and density, might be influenced by Coulomb as well as by finite size effects. For these reasons, it is highly needed to correct for these effects as best as possible.  It is the goal of this paper to propose a method to correct for Coulomb effects in the exit channel of produced charged particles. In order to support our findings, we will compare our results to the neutron case, which is of course independent, at least not directly, from the Coulomb force. Of course, neutron distributions and fluctuations are not easily determined experimentally, thus we will base our considerations on theoretical simulations using CoMD. These simulations have already been discussed in \cite{hua1, hua2, hua3} for ${}^{40}Ca+{}^{40}Ca$ at $b=1fm$ and for beam energies ranging from $4$ MeV/A to $100$ MeV/A in the laboratory system. About 250,000 events for each case have been generated, a statistics which is not sufficient in some cases as we will discuss below.
\begin{figure}
\centering
\includegraphics[width=0.5\columnwidth]{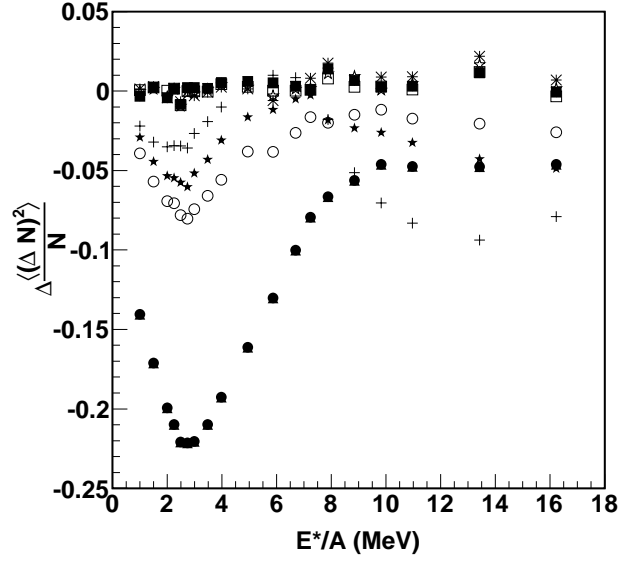}
\caption{The multiplicity fluctuation differences versus E*/A for different $p_z^{cut}$ cuts used to select particles with $-p_z^{cut}<p_z<p_z^{cut}$. $p_z^{cut}=x\times p_z^{beam}$ and $p_z^{beam}$ is the initial $p_z$ of the beam at energy $E/A$ (MeV) in the center of mass frame. Solid circles, solid up triangles, open circles, solid stars and crossings refer to  $\Delta\frac{\langle (\Delta N)^2\rangle}{N}$ for $(p, n)$ without cut, with $10p_z^{beam}$, $1p_z^{beam}$, $0.8p_z^{beam}$ and $0.5p_z^{beam}$ cuts, respectively; solid squares, solid down triangles, open squares, open stars and asterisk refer to $\Delta\frac{\langle (\Delta N)^2\rangle}{N}$ for $({}^3H, {}^3He)$ without cut, with $10p_z^{beam}$, $1p_z^{beam}$, $0.8p_z^{beam}$ and $0.5p_z^{beam}$ cuts, respectively.}
\label{f1}
\end{figure} 
Let us imagine that we have a charged particle, say a proton with charge $Z_p$, leaving a system of charge $Z_s$, mass $A$ in a volume $V$. The particle momentum is ${\bf p_i}$, and it gets accelerated by the Coulomb field to the final momentum ${\bf p_f}$. Assuming a free wave function for the particle, the Coulomb field becomes:
\begin{eqnarray}
V(q) &=& \langle \psi_f|H_{int}|\psi_i\rangle \nonumber\\
&=& \frac{Z_pe}{V}\int e^{-i{\bf p_f\cdot x}/\hbar} \phi({\bf x})e^{i {\bf p_i \cdot x}/\hbar} d^3x\nonumber\\
&=& \frac{Z_pe}{V}\int \phi({\bf x})e^{i {\bf q \cdot x}/\hbar} d^3x\nonumber\\
&=& \frac{4\pi \alpha \hbar^3Z_pZ_s }{|{\bf q}|^2 V} \int f({\bf x}) e^{i \bf q \cdot \bf x/\hbar} d^3x\nonumber\\
&=& \frac{1.44\times 4\pi  \hbar^2Z_pZ_s }{q^2 V} \int f({\bf x}) e^{i {\bf q \cdot x}/\hbar} d^3x\nonumber\\
&=& \frac{1.44\times 4\pi  \hbar^2Z_pZ_s }{q^2 V}F({\bf q}), \label{formfactor}
\end{eqnarray}
where ${\bf q=p_i-p_f}$, $\phi({\bf x})$ is the Coulomb potential of the source, $F({\bf q})$ is the form factor \cite{povh}. This is similar to the density determination of the source for instance in electron-nucleus scattering. To make calculations feasible, we will assume that ${\bf p_i}$ is negligible, which is not a bad approximation at low energies or temperatures since most of the charged particle acceleration is due to Coulomb. At high excitation energies we expect Coulomb to be negligible \cite{aldo2, huang} since the source is at low density. In fact we have seen in previous calculations \cite{hua1, hua2, hua3} that charged and uncharged particles produced in the collisions at high energies give similar values of $T$ as expected. For simplicity we will also assume that the form factor is equal to $1$. A different form is feasible but it needs the introduction of another parameter, which is connected to the density of the source. We have tried using a Gaussian density distribution of the source, but the extra parameter calls for other conditions to be implemented and to very high statistics. We are presently studying such cases. 

The reason for essentially making a Fourier transform of the Coulomb field, is because the distribution function is modified by the factor \cite{landau}:
\begin{equation}
f(p)\propto \exp[-\frac{R_{min}}{T}]\propto \exp[-\frac{V(q=p)}{T}].
\end{equation}
 \begin{figure}
\centering
\includegraphics[width=0.5\columnwidth]{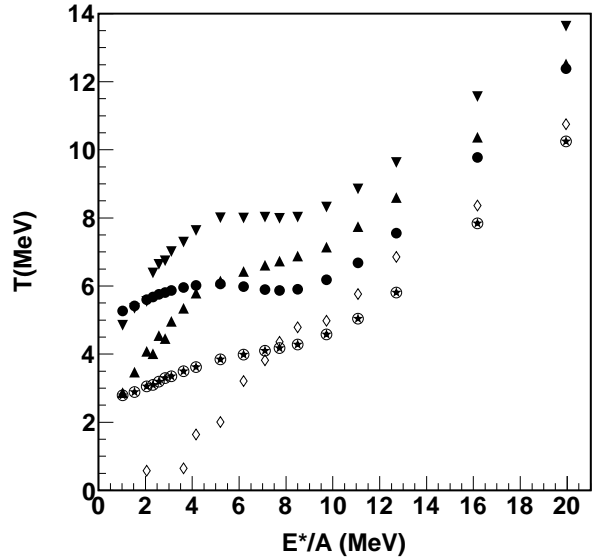}
\caption{The temperatures versus E*/A for different particles with and without Coulomb corrections. Solid circles, solid stars, solid down triangles and solid up triangles refer to $T$ of $p$, $n$, ${}^3He$ and ${}^3H$ without Coulomb corrections respectively; open circles refer to $T$ of mirror nuclei $(p, n)$ with Coulomb corrections, open diamonds refer to $T$ of mirror nuclei  $({}^3H, {}^3He)$ with Coulomb corrections.  $d$ and $\alpha$ are assumed to have the same $T$ as the neutrons and are not included in the figure for clarity.}
\label{f2}
\end{figure} 
Using this result, we can estimate modifications to physical quantities in the classical and quantum cases. The classical case is interesting because, as we will show, gives smaller temperatures for different fragments, very close to the neutron case. Furthermore, since we have an extra parameter, the volume $V$, entering Eq. (\ref{formfactor}), we need a further condition in order to determine both quantities, $V$ and $T$. Multiplicity fluctuations are equal to one in the classical case and the Coulomb correction does not change such a result significantly as we have shown in figure \ref{f1} for mirror nuclei. Thus the Coulomb correction is more important for kinetic quantities, quadrupole fluctuations, kinetic energy distributions, etc, and not for multiplicity fluctuations or yields. This remains true in the quantum case, where we will see that the temperatures say of protons are very close to those of neutrons after the Coulomb correction while their densities are practically independent on it. We stress that, in the quantum case, the density is mainly determined by the MF. In the next sections we will discuss the classical and quantum cases separately and draw the conclusions in the last section.

\begin{figure}
\centering
\includegraphics[width=0.5\columnwidth]{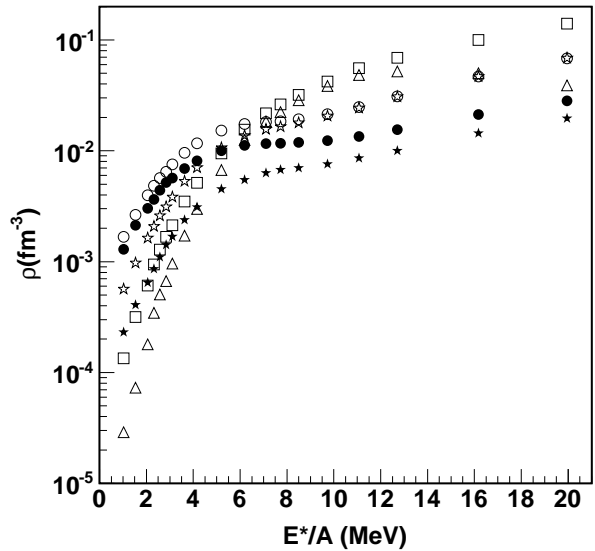}
\caption{Densities versus E*/A. Solid circles, solid stars refer to densities of $p$, $n$ obtained from quantum fluctuations without coulomb correction respectively; open circles, open stars, open squares and open triangles refer to densities of  $p$, $n$, $d$ and $\alpha$ obtained from  Eq. (\ref{rhonv}) respectively.  Notice that the high density obtained in the $d$-case is most probably due to the fact that they are overbound in the CoMD model. Experimental results display a different ordering \cite{qin} because of the different binding energies in the model.}
\label{f3}
\end{figure} 

\section{Classical case}
The quadrupole momentum in the transverse direction to the beam axis (to minimize dynamical effects) was defined in \cite{sara}
\begin{equation}
Q_{xy} = p_x^2-p_y^2.
\end{equation}
The quadrupole momentum fluctuations including the Coulomb corrections are given by:
\begin{equation}
\langle \sigma_{xy}^2 \rangle = \frac{\int d^3p (p_x^2-p_y^2)^2 e^{-(\frac{p^2}{2mT}+\frac{1.44\times 4\pi\hbar^2 Z_pZ_s }{p^2VT})}}{\int d^3p e^{-(\frac{p^2}{2mT}+ \frac{1.44\times 4\pi\hbar^2 Z_pZ_s}{p^2VT})}}.
\end{equation}
where $Z_i$ are the charges of the source and accelerated ion. After some algebra reported in appendix A we get
\begin{equation}
\langle \sigma_{xy}^2 \rangle = \frac{1}{a^2} [1+\frac{\frac{8}{5}ab+\frac{8}{15}(ab)^{3/2}}{1+2(ab)^{1/2}}], \label{classcoulcorrqf}
\end{equation}
where
\begin{equation}
a=\frac{1}{2mT}, \quad b=\frac{1.44\times 4\pi\hbar^2 Z_pZ_s }{VT}.
\end{equation}
The first term in Eq. (\ref{classcoulcorrqf}) agrees with the classical result obtained in \cite{sara} and the correction depends on the charge, volume and mass of the emitted particle and source. Within the same spirit we can calculate the multiplicity fluctuations, which we report in appendix B. The derived multiplicity fluctuations are not able to reproduce the results obtained in CoMD for $p$, $n$,  ${}^3H$ and ${}^3He$.  In particular we show in figure 1 that the multiplicity fluctuations of ${}^3H$ and ${}^3He$ are very similar, suggesting that Coulomb is not responsible for their quenching. In the same figure we display the difference of MF of protons and neutrons. Such a difference is quite large, which would suggest a Coulomb effect. However, we notice that the difference is especially large at low beam energy when the nucleons are probably emitted from the touching surfaces of the colliding nuclei. If this is true then the emitted proton or neutron might be differently reabsorbed by one of the nuclei in some sort of shadowing. Thus, if we restrict the multiplicity fluctuations of particles in the direction perpendicular to the beam axis, then their difference should be small. As we see in the figure, this is indeed the case when we calculate the MF for particles emitted with a small momentum along the beam axis, i.e. particles, which are predominantly emitted perpendicular to the beam. Notice that this strategy agrees with the choice of calculating the QF and the excitation energy \cite{hua1, hua2, hua3} in the perpendicular direction.

Since MF cannot give any further constraints in the classical case, we need a different strategy in order to solve Eq. (\ref{classcoulcorrqf}), which depends on $T$ and $V$. Let us assume that mirror nuclei, for instance ${}^3H$ and ${}^3He$, behave similarly the only differences due to the Coulomb shift in the exit channel. If this is true, then $T$ and $V$ are the same for the two particles. Thus we can write down two equations for each case and from these derive the values of $T$ and $V$.  Of course the value of $T$ will be smaller than their respective values obtained without Coulomb correction, when say ${}^3He$, displays a higher temperature than ${}^3H$. This is indeed observed in the experimental data as well \cite{sara}. In figure \ref{f2} we plot the $T$ obtained with and without Coulomb corrections for those mirror nuclei as function of the excitation energy. As predicted the Coulomb corrected temperature is smaller than the uncorrected ones. Further, their common value is very close to that obtained from the neutrons. We notice that the discrepancy observed at small excitation energies might be due mostly to the low statistics of those particles, especially ${}^3He$, in the calculations. In particular the number of points for ${}^3He$, ${}^3H$ displayed in the figure is much less than the $p$, $n$ points, because of low statistics. Adopting such a strategy we can derive the $T$ for other mirror nuclei such as $p$ and $n$. Trivially the new $T$ will coincide with the neutron one. However, in experimental data where the neutron's $T$ is not measured, one could assume that $T$ is given by the ${}^3He$, ${}^3H$ mirror nuclei and from the proton QF one could derive the $V$ which does not need to be the same as that of the other mirror nuclei \cite{kris}. The same strategy can be adopted to determine the $V$ seen by $d$ and $\alpha$ particles. All these cases are displayed in figure \ref{f2}.

In cases where high statistics is available, for instance in experiments, one could determine $T$ and $V$ from other mirror nuclei such as ${}^7Li$, ${}^7Be$ etc. and confirm if they agree or not with the previously determined ones. Our calculations do not allow us to do so because of the low statistics of those particles.
From the volume, we can calculate the density for each particle type as
\begin{equation}
\rho=\frac{N}{V},\label{rhonv}
\end{equation}
where $N$ is the multiplicity of the particle. In figure \ref{f3} we plot the density vs excitation energy per particle in different cases and we compare to the density obtained from quantum fluctuations \cite{hua1, hua2}. All results have been obtained using $1p_z^{beam}$ cut, a compromise to include particles going in the perpendicular direction and enough statistics. A dependence on the particle type is present, similar to experimental observations \cite{qin}.  We have estimated the density of $d$ and $\alpha$ as well, by assuming that they have the same neutron temperature. We stress that the assumption of equal temperature of different particles is perfectly in the spirit of an ergodic system and it is used, for instance, when calculating $T$ from the double isotope ratio \cite{albergo}.  From figure \ref{f2}, the 'near ergodicity' of the system is supported from the $T$ similarity of  neutrons with ${}^3He$, ${}^3H$. We will find a similar result in the quantum case.

\begin{figure}
\centering
\includegraphics[width=0.5\columnwidth]{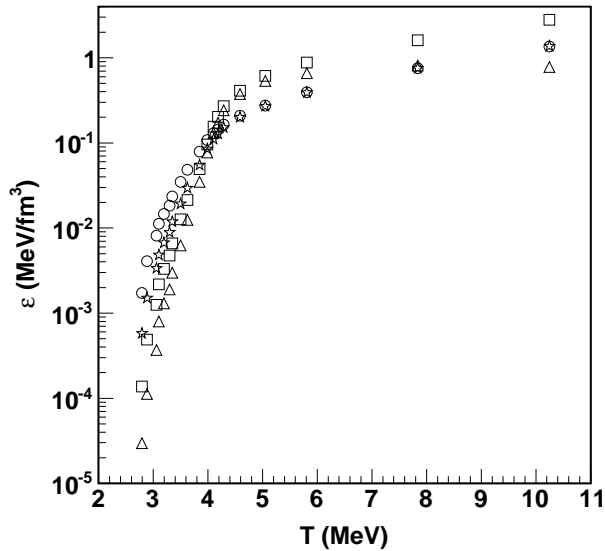}
\caption{Energy densities versus temperature. Open circles, open stars, open squares and open triangles refer to $p$, $n$, $d$ and $\alpha$ obtained from the classical case with Coulomb correction.}
\label{f4}
\end{figure}

From the values of density and excitation energy, we can easily obtain the energy density, which is plotted in figure \ref{f4} as function of $T$. The plot displays the same features reported in \cite{hua1, hua2, hua3}. In particular the very rapid increase at small $T$ is due to the opening of many evaporation channels which terminates around $T=4MeV$ when fragmentation starts. The fragmentation region terminates around $T=10MeV$ for $p$ and $n$, close to the critical temperature \cite{justin}. Quantum corrections, as we will discuss in the next section, gives qualitatively similar results.

\section{Quantum case}
The above discussion can be generalized to the quantum case. In particular, in this work we will restrict the results to the $p$ and $n$ cases (fermions) and avoid involved discussions on bosons ($d$ and $\alpha$) or more complex Fermions. We are currently working on those cases.
The QF can be obtained from:
\begin{equation}
\langle \sigma_{xy}^2 \rangle=(2mT)^2\frac{4}{15}\frac{\int_0^\infty dy y^\frac{5}{2} \frac{1}{e^{y+\frac{A}{yT^2}-\nu}+1}}{\int_0^\infty dy y^\frac{1}{2} \frac{1}{e^{y+\frac{A}{yT^2}-\nu}+1}},
\label{quadcoulcorrqf}
\end{equation}
where $A=\frac{1.44\times4\pi\hbar^2 q_1q_2}{2mV}$ and $\nu=\frac{\mu}{T}$. The terms in Eq. (\ref{quadcoulcorrqf}) are similar to their classical counterpart and a detailed derivation of this result is given in appendix C. On the same ground we can derive the MF as:
\begin{equation}
\frac{\langle(\Delta N)^2\rangle}{N}=\frac{\int_0^\infty dy y^\frac{1}{2}\frac{e^{y+\frac{A}{yT^2}-\nu}}{(e^{y+\frac{A}{yT^2}-\nu}+1)^2}}{\int_0^\infty dy y^\frac{1}{2}\frac{1}{e^{y+\frac{A}{yT^2}-\nu}+1}}.
\label{quancoulcorrmf}
\end{equation}

\begin{figure}
\centering
\includegraphics[width=0.5\columnwidth]{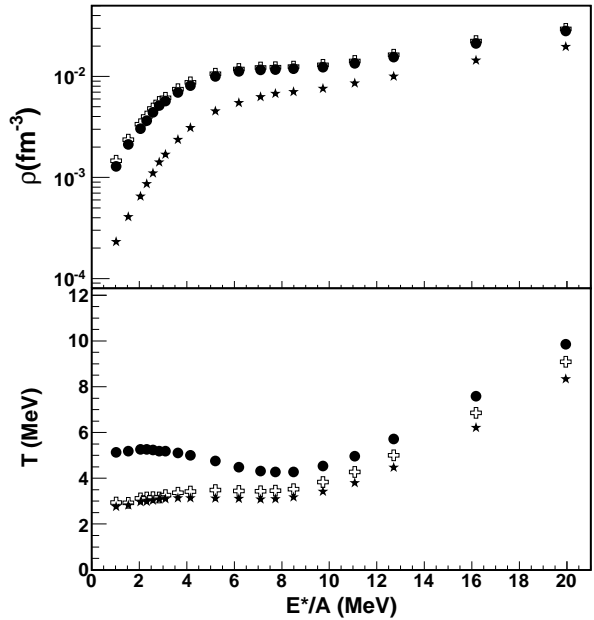}
\caption{(Top panel) Densities versus E*/A. (Bottom panel) Temperatures versus E*/A. Solid circles and solid stars refer to $p$ and $n$ obtained from quantum fluctuations without Coulomb correction respectively; open crosses refer to $p$-case obtained from quantum fluctuations with Coulomb correction.}
\label{f5}
\end{figure} 
Again the detailed derivation is given in the appendix C. Those equations can be solved numerically. In figure \ref{f5} we plot $T$ and $\rho$ vs excitation energy respectively. The protons and neutrons cases only are included. As we see the derived $T$ of protons are much closer to the neutrons, supporting the ansatz we used in the classical case. Also the good agreement for the obtained temperatures suggests that thermal equilibrium in the transverse direction is nearly reached.The modification to the density due to Coulomb is very small which implies that the MF are not so much affected by Coulomb. 
\begin{figure}
\centering
\includegraphics[width=0.5\columnwidth]{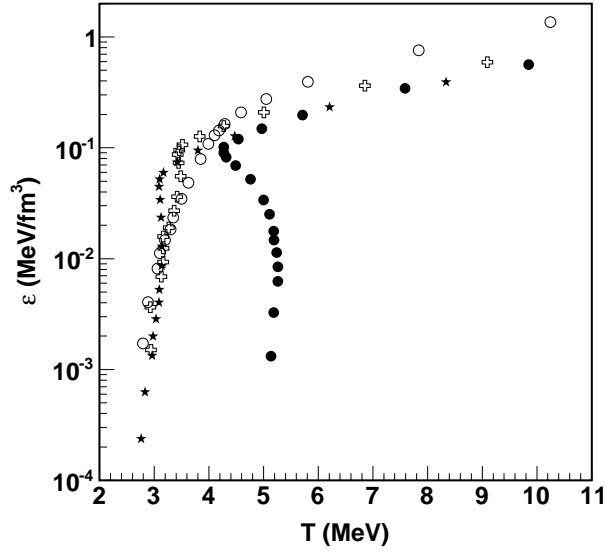}
\caption{Energy density versus temperature. Open circles refer to the classical results with Coulomb correction, other symbols as in figure \ref{f5}.}
\label{f6}
\end{figure} 

As we see from the results, even though the $T$ are similar for $p$ and $n$, their densities are not which suggests that p and n 'see' different densities probably already in the ground state of the nuclei. Those differences are less noticeable if we plot the energy density $\varepsilon=\frac{E}{N}\rho$ versus $T$. This is displayed in figure \ref{f6}, which shows a very similar behavior of $p$ and $n$. 


\section{Conclusion}
In conclusion, in this paper we have discussed Coulomb modifications to the density and temperature in heavy ion collisions. The classical and quantum cases (fermions only) have been discussed. We have shown that in both cases, the temperatures obtained from different particle types are very similar to the neutron's one which implies the 'near ergodicity' of the system. On the other hand the densities are different for different particles, which suggests that the Coulomb dynamics is of course important also before the breaking of the source. The energy densities are very similar at high temperatures, which implies that Coulomb corrections are small due to the small source densities. Experimental investigations of the effects discussed in this work for well determined sources and excitation energies \cite{alan1, sara, justin, qin, borderie} would be very important to further constrain the Nuclear Equation of State in the liquid-gas phase transition region also for asymmetric matter. The role of pairing and the possibility of a Bose condensate should also be further investigated. 

A mathematica code for the quantum case is available from the authors upon request.

\appendix
\section{}
For the classical case, assuming particles follow the Maxwell-Boltzmann distribution, then the momentum quadrupole fluctuation including the Coulomb effect is: 
\begin{eqnarray}
\langle \sigma_{xy}^2 \rangle&=&\frac{\int d^3p (p_x^2-p_y^2)^2 e^{-(\frac{p^2}{2mT}+\frac{1.44\times 4\pi\hbar^2 Z_pZ_s }{p^2VT})}}{\int d^3p e^{-(\frac{p^2}{2mT}+ \frac{1.44\times 4\pi\hbar^2 Z_pZ_s}{p^2VT})}}\nonumber\\
&=&\frac{\int d^3p (p_x^2-p_y^2)^2 e^{-(ap^2+\frac{b }{p^2})}}{\int d^3p e^{-(ap^2+\frac{b }{p^2})}}.\label{quadrupole1}
\end{eqnarray}
For simplicity, we write
\begin{equation}
a=\frac{1}{2mT}, \quad b=\frac{1.44\times 4\pi\hbar^2 Z_pZ_s }{VT}.\label{coe1}
\end{equation}
Using 
\begin{equation}
p_x=p\sin\theta \cos\phi, \quad p_y=p\sin\theta \sin\phi, \quad p_z=p\cos\theta,
\end{equation}
one obtains
\begin{eqnarray}
\langle \sigma_{xy}^2 \rangle&=&\frac{\int d^3p (p_x^2-p_y^2)^2 e^{-(ap^2+\frac{b }{p^2})}}{\int d^3p e^{-(ap^2+\frac{b }{p^2})}}\nonumber\\
&=&\frac{\int_0^\infty dp p^6 e^{-(ap^2+\frac{b }{p^2})}\int_0^{\pi}d\theta \sin^5\theta\int_0^{2\pi}d\phi(\cos^2\phi-\sin^2\phi)^2}{\int_0^\infty dp p^2 e^{-(ap^2+\frac{b }{p^2})}\int_0^{\pi}d\theta \sin\theta\int_0^{2\pi}d\phi}\nonumber\\
&=&\frac{4}{15} \frac{\int_0^\infty dp p^6 e^{-(ap^2+\frac{b }{p^2})}}{\int_0^\infty dp p^2 e^{-(ap^2+\frac{b }{p^2})}}.\label{quadrupole2}
\end{eqnarray}
Define the integral 
\begin{equation}
I_n = \int_0^\infty dx x^n e^{-(ax^2+\frac{b }{x^2})},
\end{equation}
where $a>0, b>0$. Then
\begin{equation}
\langle \sigma_{xy}^2 \rangle = \frac{4}{15}\frac{I_6}{I_2}. \label{quadrupole3}
\end{equation}
Now we are going to calculate the integral $I_n$,
\begin{eqnarray}
I_n &=&  \int_0^\infty dx x^n e^{-(ax^2+\frac{b }{x^2})}\nonumber\\
&=& \frac{1}{n+1} \int_0^\infty dx^{n+1} e^{-(ax^2+\frac{b }{x^2})}\nonumber\\
&=& -\frac{1}{n+1} \int_0^\infty dx x^{n+1}[-(2ax-\frac{2b }{x^3})]e^{-(ax^2+\frac{b }{x^2})}\nonumber\\
&=& \frac{2}{n+1} \int_0^\infty dx x^{n+1}(ax-\frac{b}{x^3})e^{-(ax^2+\frac{b }{x^2})}\nonumber\\
&=& \frac{2}{n+1} [a \int_0^\infty dx x^{n+2}e^{-(ax^2+\frac{b }{x^2})}-b\int_0^\infty dx x^{n-2}e^{-(ax^2+\frac{b }{x^2})}]\nonumber\\
&=& \frac{2}{n+1} [a I_{n+2}-b I_{n-2}]. \label{in2}
\end{eqnarray}
Then
\begin{equation}
I_{n+2} = \frac{n+1}{2a}I_n+\frac{b}{a}I_{n-2}.\label{in}
\end{equation}
We derived the recurrence relation for the integral $I_n$.  If we know two of them, we can calculate all the integrals. On the other hand, 
\begin{eqnarray}
I_n &=&  \int_0^\infty dx x^n e^{-(ax^2+\frac{b }{x^2})}\nonumber\\
&=& \int_0^\infty dx x^n e^{-(ab)^{1/2}[(\frac{a}{b})^{1/2}x^2+\frac{1}{(\frac{a}{b})^{1/2}x^2}]}\nonumber\\
&=& (\frac{b}{a})^{(n+1)/4} \int_0^\infty dy y^n e^{-(ab)^{1/2}[y^2+\frac{1}{y^2}]}\nonumber\\
&=& (\frac{b}{a})^{(n+1)/4}e^{-2(ab)^{1/2}} \int_0^\infty dy y^n e^{-(ab)^{1/2}(y-\frac{1}{y})^2}. \label{in1}
\end{eqnarray}
First we calculate $I_0$. Let $n=0$ in Eq. (\ref{in1})
\begin{eqnarray}
I_0 &=&  (\frac{b}{a})^{1/4}e^{-2(ab)^{1/2}} \int_0^\infty dy e^{-(ab)^{1/2}(y-\frac{1}{y})^2} \nonumber\\
&=& \frac{1}{2}(\frac{b}{a})^{1/4}e^{-2(ab)^{1/2}} [\int_0^\infty dy e^{-(ab)^{1/2}(y-\frac{1}{y})^2}+\int_0^\infty dx \frac{1}{x^2} e^{-(ab)^{1/2}(x-\frac{1}{x})^2}]\nonumber\\
&=& \frac{1}{2}(\frac{b}{a})^{1/4}e^{-2(ab)^{1/2}} \int_0^\infty dy (1+\frac{1}{y^2}) e^{-(ab)^{1/2}(y-\frac{1}{y})^2}\nonumber\\
&=&  \frac{1}{2}(\frac{b}{a})^{1/4}e^{-2(ab)^{1/2}}\int_0^\infty d(y-\frac{1}{y}) e^{-(ab)^{1/2}(y-\frac{1}{y})^2}\nonumber\\
&=& \frac{1}{2}(\frac{b}{a})^{1/4}e^{-2(ab)^{1/2}}\int_{-\infty}^\infty dx e^{-(ab)^{1/2}x^2}\nonumber\\
&=& \frac{\pi^{1/2}}{2 a^{1/2}}e^{-2(ab)^{1/2}}.\label{i0}
\end{eqnarray}
Second we calculate $I_{-2}$. Let $n=-2$ in Eq. (\ref{in1})
\begin{eqnarray}
I_{-2} &=&  (\frac{b}{a})^{(-2+1)/4}e^{-2(ab)^{1/2}} \int_0^\infty dy y^{-2}e^{-(ab)^{1/2}(y-\frac{1}{y})^2} \nonumber\\
&=& (\frac{b}{a})^{-1/2}(\frac{b}{a})^{1/4}e^{-2(ab)^{1/2}} \int_0^\infty dy \frac{1}{y^2}e^{-(ab)^{1/2}(y-\frac{1}{y})^2} \nonumber\\
&=& (\frac{b}{a})^{-1/2}(\frac{b}{a})^{1/4}e^{-2(ab)^{1/2}} \int_0^\infty dx e^{-(ab)^{1/2}(x-\frac{1}{x})^2} \nonumber\\
&=&(\frac{b}{a})^{-1/2}I_0\nonumber\\
&=&(\frac{b}{a})^{-1/2} \frac{\pi^{1/2}}{2 a^{1/2}}e^{-2(ab)^{1/2}}\nonumber\\
&=& \frac{\pi^{1/2}}{2 b^{1/2}}e^{-2(ab)^{1/2}}. \label{i-2}
\end{eqnarray}
Using Eqs. (\ref{in}, \ref{i0}, \ref{i-2}), we can calculate
\begin{eqnarray}
I_2 &=&  \frac{1}{2a}I_0+\frac{b}{a}I_{-2}\nonumber\\
&=&  \frac{1}{2a}I_0+\frac{b}{a}(\frac{b}{a})^{-1/2} I_0\nonumber\\
&=& [\frac{1}{2a}+(\frac{b}{a})^{1/2}] I_0 \nonumber\\
&=&  [\frac{1}{2a}+(\frac{b}{a})^{1/2}]\frac{\pi^{1/2}}{2 a^{1/2}}e^{-2(ab)^{1/2}} \nonumber\\
&=& \frac{1+2(ab)^{1/2}}{4a^{3/2}}\pi^{1/2}e^{-2(ab)^{1/2}}. \label{i2}
\end{eqnarray}
\begin{eqnarray}
I_4 &=&  \frac{3}{2a}I_2+\frac{b}{a}I_{0}\nonumber\\
&=& \frac{3}{2a}[\frac{1}{2a}+(\frac{b}{a})^{1/2}] I_0+\frac{b}{a}I_{0}\nonumber\\
&=& [\frac{3}{(2a)^2}+\frac{3}{2a}(\frac{b}{a})^{1/2}+\frac{b}{a}]I_0\nonumber\\
&=& [\frac{3}{(2a)^2}+\frac{3}{2a}(\frac{b}{a})^{1/2}+\frac{b}{a}]\frac{\pi^{1/2}}{2 a^{1/2}}e^{-2(ab)^{1/2}}\nonumber\\
&=& \frac{3+6(ab)^{1/2}+4ab}{8a^{5/2}}\pi^{1/2}e^{-2(ab)^{1/2}}. \label{i4}
\end{eqnarray}
\begin{eqnarray}
I_6 &=&  \frac{5}{2a}I_4+\frac{b}{a}I_{2}\nonumber\\
&=& \frac{5}{2a}[\frac{3}{(2a)^2}+\frac{3}{2a}(\frac{b}{a})^{1/2}+\frac{b}{a}]I_0+\frac{b}{a} [\frac{1}{2a}+(\frac{b}{a})^{1/2}] I_0\nonumber\\
&=& [\frac{15}{(2a)^3}+\frac{15}{(2a)^2}(\frac{b}{a})^{1/2}+\frac{3b}{a^2}+(\frac{b}{a})^{3/2}] I_0 \nonumber\\
&=&  [\frac{15}{(2a)^3}+\frac{15}{(2a)^2}(\frac{b}{a})^{1/2}+\frac{3b}{a^2}+(\frac{b}{a})^{3/2}] \frac{\pi^{1/2}}{2 a^{1/2}}e^{-2(ab)^{1/2}}\nonumber\\
&=& \frac{15+30(ab)^{1/2}+24ab+8(ab)^{3/2}}{16a^{7/2}}\pi^{1/2}e^{-2(ab)^{1/2}}. \label{i6}
\end{eqnarray}
Substitute Eqs. (\ref{i2}, \ref{i6}) into Eq. (\ref{quadrupole3}), we obtain
\begin{eqnarray}
\langle \sigma_{xy}^2 \rangle &=& \frac{4}{15}\frac{I_6}{I_2}\nonumber\\
&=&\frac{4}{15}\frac{\frac{15+30(ab)^{1/2}+24ab+8(ab)^{3/2}}{16a^{7/2}}\pi^{1/2}e^{-2(ab)^{1/2}}}{ \frac{1+2(ab)^{1/2}}{4a^{3/2}}\pi^{1/2}e^{-2(ab)^{1/2}}}\nonumber\\
&=&\frac{4}{15} \frac{1}{4a^2} \frac{15+30(ab)^{1/2}+24ab+8(ab)^{3/2}}{1+2(ab)^{1/2}}\nonumber\\
&=& \frac{1}{a^2} \frac{1+2(ab)^{1/2}+\frac{8}{5}ab+\frac{8}{15}(ab)^{3/2}}{1+2(ab)^{1/2}}\nonumber\\
&=&\frac{1}{a^2} [1+\frac{\frac{8}{5}ab+\frac{8}{15}(ab)^{3/2}}{1+2(ab)^{1/2}}].
\end{eqnarray}

\section{}
For the classical case, the single particle partition function considering the Coulomb effect is
\begin{eqnarray}
Z_1 &=& \frac{1}{h^3}\int e^{-\beta \varepsilon} d^3x d^3p\nonumber\\
&=& \frac{4\pi V}{h^3}\int_0^\infty e^{-(\frac{p^2}{2mT}+\frac{1.44\times 4\pi Z_pZ_s}{VT p^2})}p^2 dp \nonumber\\
&=& \frac{4\pi V}{h^3}\int_0^\infty e^{-(a p^2+\frac{b}{p^2})}p^2 dp\nonumber\\
&=&  \frac{4\pi V}{h^3} \times[\frac{1+2(ab)^{1/2}}{4a^{3/2}}\pi^{1/2} e^{-2(ab)^{1/2}}].
\end{eqnarray}
Then the pressure is
\begin{eqnarray}
P &=& \frac{N}{\beta}\frac{\partial }{\partial V} \ln Z_1\nonumber\\
&=&NT \frac{\partial }{\partial V} \ln \{\frac{4\pi V}{h^3} \times[\frac{1+2(ab)^{1/2}}{4a^{3/2}}\pi^{1/2} e^{-2(ab)^{1/2}}]\}\nonumber\\
&=& NT \frac{\partial }{\partial V} \ln \{V \times[1+2(ab)^{1/2}]e^{-2(ab)^{1/2}}\}\nonumber\\
&=& NT \frac{\partial }{\partial V} \ln \{V \times[1+2(ab')^{1/2}V^{-1/2}]e^{-2(ab')^{1/2}V^{-1/2}}\}\nonumber\\
&=&NT[\frac{1}{V}-\frac{(ab')^{1/2}V^{-3/2}}{1+2(ab')^{1/2}V^{-1/2}}+(ab')^{1/2}V^{-3/2} ] \nonumber\\
&=&NT[\frac{1}{V}+\frac{2ab'V^{-2}}{1+2(ab')^{1/2}V^{-1/2}} ],
\end{eqnarray}
where $b'=\frac{1.44\times 4\pi \hbar^2 Z_pZ_s}{T}$. Thus
\begin{eqnarray}
\frac{\partial P}{\partial V}|_{N, T} &=&NT\{-\frac{1}{V^2}+\frac{-4ab'V^{-3}[1+2(ab')^{1/2}V^{-1/2}]-2ab'V^{-2}[-(ab')^{1/2}V^{-3/2}]}{[1+2(ab')^{1/2}V^{-1/2}]^2} \}\nonumber\\
&=&NT\{-\frac{1}{V^2}-\frac{4ab'}{[1+2(ab')^{1/2}V^{-1/2}]V^3}+\frac{2(ab')^{3/2}}{[1+2(ab')^{1/2}V^{-1/2}]^2V^{7/2}} \}\nonumber\\
&=&-\frac{NT}{V^2}\{1+\frac{4ab'}{[1+2(ab')^{1/2}V^{-1/2}]V}-\frac{2(ab')^{3/2}}{[1+2(ab')^{1/2}V^{-1/2}]^2V^{3/2}} \}.
\end{eqnarray}
The normalized multiplicity fluctuation is 
\begin{eqnarray}
\frac{\langle (\Delta N )^2 \rangle}{N}&=&-\frac{TN}{V^2}\frac{\partial V}{\partial P}|_{T, N}\nonumber\\
&=&-\frac{TN}{V^2}\times \frac{1}{-\frac{NT}{V^2}\{1+\frac{4ab'}{[1+2(ab')^{1/2}V^{-1/2}]V}-\frac{2(ab')^{3/2}}{[1+2(ab')^{1/2}V^{-1/2}]^2V^{3/2}} \}}\nonumber\\
&=&\frac{1}{1+\frac{4ab'}{[1+2(ab')^{1/2}V^{-1/2}]V}-\frac{2(ab')^{3/2}}{[1+2(ab')^{1/2}V^{-1/2}]^2V^{3/2}} }.
\end{eqnarray}
To simplify the above equation, we define 
\begin{equation}
x=\frac{ab'}{V}, \label{b5}
\end{equation}
then
\begin{equation}
\frac{\langle (\Delta N )^2 \rangle}{N}=\frac{1}{1+\frac{4x}{1+2x^{1/2}}-\frac{2x^{3/2}}{(1+2x^{1/2})^2} }. \label{b6}
\end{equation}
The last equation (\ref{b6}) cannot be directly applied to the multiplicity fluctuations say of protons, since we know most of those fluctuations are due to Fermion quenching. In fact the protons and neutrons multiplicity fluctuations are very similar when observed in the perpendicular direction to the beam, see figure \ref{f1}. In practice one could apply Eq. (\ref{b6}) to the difference between $p$ and $n$ or ${}^3He$, ${}^3H$ multiplicity fluctuations which we could not do because of low statistics in the model case.
\section{}
For the quantum case, assuming particles follow the Fermi-Dirac distribution,
\begin{equation}
f(p)=\frac{1}{e^{[\varepsilon+\frac{1.44\times4\pi\hbar^2 Z_pZ_s}{Vp^2}-\mu]/T}+1},
\end{equation}
where $\varepsilon=\frac{p^2}{2m}$ is the energy , $\mu$ is the chemical potential, $T$ is the temperature.
The average number of particles is
\begin{eqnarray}
N &=&\frac{g}{h^3}\int d^3xd^3p f(p)\nonumber\\
&=&\frac{gV}{h^3}4\pi\int_0^\infty dpp^2 f(p) .
\label{av_particle1}
\end{eqnarray}
Let's make the integral variable transformation,
\begin{equation}
\varepsilon=\frac{p^2}{2m},\quad p=(2m\varepsilon)^{\frac{1}{2}}, \quad dp=\frac{m}{\sqrt{2m\varepsilon}}d\varepsilon. \label{transformation1}
\end{equation}
Thus Eq. (\ref{av_particle1}) becomes
\begin{eqnarray}
N &=&\frac{gV}{h^3}4\pi\int_0^\infty dpp^2f(p)\nonumber\\
&=&\frac{gV}{h^3}4\pi\frac{(2m)^\frac{3}{2}}{2}\int_0^\infty d\varepsilon \varepsilon^\frac{1}{2} f(\varepsilon)\nonumber\\
&=&\frac{gV}{h^3}4\pi\frac{(2m)^\frac{3}{2}}{2}\int_0^\infty d\varepsilon \varepsilon^\frac{1}{2}\frac{1}{e^{[\varepsilon+ \frac{1.44\times4\pi\hbar^2 Z_pZ_s}{Vp^2}-\mu]/T}+1}\nonumber\\
&=&\frac{gV}{h^3}4\pi\frac{(2m)^\frac{3}{2}}{2}\int_0^\infty d\varepsilon \varepsilon^\frac{1}{2}\frac{1}{e^{[\varepsilon+ \frac{1.44\times4\pi\hbar^2 Z_pZ_s}{2mV\varepsilon}-\mu]/T}+1}\nonumber\\
&=&\frac{gV}{h^3}4\pi\frac{(2m)^\frac{3}{2}}{2}\int_0^\infty d\varepsilon \varepsilon^\frac{1}{2}\frac{1}{e^{[\varepsilon+ \frac{A}{\varepsilon}-\mu]/T}+1},
\label{av_particle2}
\end{eqnarray}
where $A=\frac{1.44\times4\pi\hbar^2 Z_pZ_s}{2mV}$. Let's make the integral variable transformation again
\begin{equation}
y=\frac{\varepsilon}{T}, \quad \nu=\frac{\mu}{T}.\label{notation}
\end{equation}
Therefore, Eq. (\ref{av_particle2}) becomes
\begin{equation}
N =\frac{gV}{h^3}4\pi\frac{(2mT)^\frac{3}{2}}{2}\int_0^\infty dy y^\frac{1}{2}\frac{1}{e^{y+\frac{A}{yT^2}-\nu}+1}.
\label{av_particle3}
\end{equation}
The multiplicity fluctuation is
\begin{equation}
\langle(\Delta N)^2\rangle=T(\frac{\partial N}{\partial \mu})_{T, V}=(\frac{\partial N}{\partial \nu})_{T, V}.\label{fluctuation1}
\end{equation}
Substitute Eq. (\ref{av_particle3}) into Eq. (\ref{fluctuation1}), one can obtain
\begin{eqnarray}
\langle(\Delta N)^2\rangle=\frac{gV}{h^3}4\pi\frac{(2mT)^\frac{3}{2}}{2}\int_0^\infty dy y^\frac{1}{2}\frac{e^{y+\frac{A}{yT^2}-\nu}}{(e^{y+\frac{A}{yT^2}-\nu}+1)^2}.\label{fluctuation3}
\end{eqnarray}
Divide Eq. (\ref{fluctuation3}) by Eq. (\ref{av_particle3}), one can get
\begin{eqnarray}
\frac{\langle(\Delta N)^2\rangle}{N}&=&\frac{\frac{gV}{h^3}4\pi\frac{(2mT)^\frac{3}{2}}{2}\int_0^\infty dy y^\frac{1}{2}\frac{e^{y+\frac{A}{yT^2}-\nu}}{(e^{y+\frac{A}{yT^2}-\nu}+1)^2}}{\frac{gV}{h^3}4\pi\frac{(2mT)^\frac{3}{2}}{2}\int_0^\infty dy y^\frac{1}{2}\frac{1}{e^{y+\frac{A}{yT^2}-\nu}+1}}\nonumber\\
&=&\frac{\int_0^\infty dy y^\frac{1}{2}\frac{e^{y+\frac{A}{yT^2}-\nu}}{(e^{y+\frac{A}{yT^2}-\nu}+1)^2}}{\int_0^\infty dy y^\frac{1}{2}\frac{1}{e^{y+\frac{A}{yT^2}-\nu}+1}}.
\label{fluctuation4}
\end{eqnarray}
In the same framework, we also calculate the quadrupole momentum fluctuation 
\begin{eqnarray}
\langle \sigma_{xy}^2 \rangle&=&\frac{\int d^3p (p_x^2-p_y^2)^2 \frac{1}{e^{[\frac{p^2}{2m}+\frac{1.44\times 4\pi\hbar^2 Z_pZ_s }{p^2V}-\mu]/T}+1}}{\int d^3p \frac{1}{e^{[\frac{p^2}{2m}+\frac{1.44\times 4\pi\hbar^2 Z_pZ_s }{p^2V}-\mu]/T}+1}}\nonumber\\
&=&(2mT)^2\frac{4}{15}\frac{\int_0^\infty dy y^\frac{5}{2} \frac{1}{e^{y+\frac{A}{yT^2}-\nu}+1}}{\int_0^\infty dy y^\frac{1}{2} \frac{1}{e^{y+\frac{A}{yT^2}-\nu}+1}}
\label{quadrupole4}
\end{eqnarray}


\end{document}